\documentclass[pra,twocolumn,superscriptaddress,floatfix]{revtex4}%
\usepackage{graphicx}
\usepackage{bm, amsmath, amssymb}
\usepackage{color}
\usepackage{soul}
\usepackage{pdfpages}
\usepackage{amstext}
\usepackage{amsmath}
\usepackage{amsfonts}
\usepackage{amssymb}
\usepackage{algorithm}
\usepackage{algpseudocode}
\usepackage{amsmath}%
\usepackage{appendix}

\begin{document}

\title{Optimizing quantum control pulses with complex constraints and few
variables through Tensorflow}
\author{Yao Song}
\affiliation{Shenzhen Institute of Quantum Science and Engineering, Southern University
of Science and Technology, Shenzhen, Guangdong 518055, China}

\author{Junning Li}
\affiliation{Department of Physics, City University of Hong Kong, Tat Chee Avenue,
Kowloon, Hong Kong SAR, China}
\author{Yong-Ju Hai}
\affiliation{Shenzhen Institute of Quantum Science and Engineering, Southern University
of Science and Technology, Shenzhen, Guangdong 518055, China}
\affiliation{Department of Physics, Southern University of Science and Technology,
Shenzhen 518055, China}
\author{Qihao Guo}
\affiliation{Shenzhen Institute of Quantum Science and Engineering, Southern University
of Science and Technology, Shenzhen, Guangdong 518055, China}

\author{Xiu-Hao Deng}
\email{dengxh@sustech.edu.cn}
\affiliation{Shenzhen Institute of Quantum Science and Engineering, Southern University
of Science and Technology, Shenzhen, Guangdong 518055, China}
\affiliation{Guangdong Provincial Key Laboratory of Quantum Science and Engineering
Southern University of Science and Technology, Shenzhen, Guangdong 518055,
China}

\begin{abstract}
Applying optimal control algorithms on realistic quantum systems confronts two key challenges: to efficiently adopt physical constraints in the optimization and to minimize the variables for the convenience of experimental tune-ups. In order to resolve these issues, we propose a novel algorithm by incorporating multiple constraints into the gradient optimization over piece-wise pulse constant values, which are transformed to contained numbers of the finite Fourier basis for bandwidth control. Such complex constraints and variable transformation involved in the optimization introduce extreme difficulty in calculating gradients. We resolve this issue efficiently utilizing auto-differentiation on Tensorflow. We test our algorithm by finding smooth control pulses to implement single-qubit and two-qubit gates for superconducting transmon qubits with always-on interaction, which remains a challenge of quantum control in various qubit systems. Our algorithm provides a promising optimal quantum control approach that is friendly to complex and optional physical constraints. 
\end{abstract}

\maketitle

\section{Introduction}

\label{section1}

Potential ground-breaking quantum technologies, such as quantum computing,
quantum sensing, and quantum metrology~\cite%
{quantumcomputing,quantumsensing,quantummetrology}, become more and more
feasible with the tremendous progress of quantum control technology on
various physical systems \cite{naranjo2018dynamics, basset2021quantum,
rembold2020introduction, chen2021exponential}. Based on growing knowledge
quantum systems interacting with the enviroment, multifarious approaches
have been developed to improve the control precision~\cite%
{DRAG,swift1,swift2,CRAB2,deng2021correcting,long2021universal,GOAT}.
Eventhough, further optimizing quantum control still highly relies on
numerical approaches~\cite{GRAPE1,GRAPE2,Krotov,DE, CRAB1, GOAT}. Practical
quantum optimal control (QOC)~\cite%
{ohtsuki2004generalized,lloyd2014information,ohtsuki2008monotonically,sundermann1999extensions}
should satisfy the requirements and constraints in the physical systems,
such as more realistic Hamiltonian, maximal field strength, finite sampling
rate, limited bandwidth \cite%
{jerger2019situ,hincks2015controlling,rol2020time}, etc. Also, for efficient
calibration in experiments, the control field waveform should depend on as
few variables as possible. Other than the physical considerations, the
optimization algorithm ought to be fast and accurate and should be
extensible to larger systems. As one of the most successful numerical
optimization algorithms, GRAPE has been applied to many physical systems,
including NMR qubits~\cite{GRAPE1, tovsner2009optimal, jones2010quantum,
ryan2009randomized, nielsen2007optimal}, superconducting qubits in 3D cavity~%
\cite{schutjens2013single, blais2020quantum, allen2017optimal,
blais2007quantum,blais2020circuit}, nitrogen-vacancy (NV) centers in diamond~%
\cite{niemeyer2013broadband, wang2011time, rembold2020introduction,
platzer2010optimal}, etc. However, adapting GRAPE to multiple realistic
constraints remains challenging. Different from GRAPE, another numerical
algorithm, CRAB has been proposed to generate smooth control waveforms,
towards application in cold atoms~\cite{crabincoldatom}. There are also many
other proposed algorithms leading to promising applications \cite{DE,Krotov, niu2019universal, shu2016}

Depending on how to parametrize the control field's variables, QOC
algorithms could be assorted into two classes: 1. \textit{Piecewise constant}
(PWC) discretizes the field pulse as a sequence of piecewise-constant field
strength and the value of each piece as the optimization variable~\cite%
{GRAPE2, GRAPE1, RL, kirchhoff2018optimized, larocca2020k}. 
Varying each variable causing local variation of the whole
waveform hence local variation of the expectation function, which preserves
the convexity of optimization landscape in this parameter space. And then
gradient-based optimization could efficiently proceed \cite%
{GRAPE1,Krotov,DE,RL}. A trade-off arises between inaccurate PWC dynamics
for finite discretization rate and the cost of computing the gradients
versus massive variables. Also, PWC waveforms are not smooth and could
easily contain fast fluctuations. Filtering the optimized waveform in the
post-optimization deforms the output pulses from optimum. 2. \textit{Chopped
basis} (CB) optimization uses parameters related to a finite set of basis
expanding the control waveform. The basis are usually analytic functions
such as Gaussian, tangential, or Fourier functions~\cite{CRAB1,GOAT}, etc,
which easily guarantees the smoothness of the optimized waveform. However,
adding the physical constraints deforms the processed pulses in an unknown
way\cite{CRAB1,GOAT}, so the output pulses lose the analyticity.
Furthermore, small change of one expansion coefficient leads to global
deformation of the pulse waveform, and also coefficients of different bases
affect differently. Therefore, the CB optimization landscape becomes
non-convex. Alternatively, constraints could be incorporated into
optimization with Lagrange multipliers but the calculation of the analytical
differentiation brings in another source of complexity and difficulty \cite{plick2015physical, moore2012exploring, shu2016}.

Here in this article, we propose a novel optimization algorithm, the \textit{%
Complex and Optional Constraints Optimization with Auto-differentiation}
(COCOA), to tackle the problems summarized above. The algorithm parametrizes
the control waveform in the basis of truncated Fourier series, while the
optimization performs the gradient in the convex landscape of PWC parameter
space. The transformation between the aforementioned two parametrization
systems is bridged via approximate discrete Fourier transformation (DFT) and
inverse discrete Fourier transformation (iDFT). The advantage of combining
these two parametrization systems is that all the pulse constraints can be
incorporated in the optimization process instead of post-optimization. The
application of DFT and iDFT within the optimization iteration introduces
further physical advantages: 1. Hard bounds on the bandwidth by limiting the
Fourier basis; 2. Definite expansion with pre-selected basis; 3. Analytical
expression of output pulses with a small number of parameters for tuning. 4.
Convenient application of pulse pre-distortion because the transfer/filter
functions are in the frequency domain\cite{kelly2014optimal, rol2020time,
rol2019fast}. However, embedding these constraints and the transformation of
parametrization systems into the optimization iteration raises extreme
difficulty in solving the analytical gradient equations. We resolve this
issue using auto-differentiation during one-round back-propagation process
provided by Tensorflow\cite{tensorflow2015-whitepaper}. With Tensorflow, the
automatical calculation of gradient could be further extended to various
physical systems. This paper explains in detail how COCOA works and presents
some numerical results to demonstrate the efficiency and its advantage compared with GRAPE and CRAB.

This paper is organised as follows. First, we take an overview of quantum
optimal control theories and numerical methods. In Sec.~\ref{section3} we
present the algorithmic description of COCOA and explain how it combines the advantages of two
categories QOC approaches and perform an optimization task under complex and optional constraints.
Then in Sec.~\ref{section4}, we apply COCOA to find optimal quantum-gate pulses for superconducting qubits with always-on interactions. We simulate two models: cavity mediated two-qubit system and direct capacitively-coupled systems. We demonstrate COCOA's advantages by comparing with the representative algorithms GRAPE and CRAB of the two QOC categories. In Sec.~\ref{section5}, we conclude this paper. 

\section{Quantum Optimal Control}

\label{section2}

\subsection{General formalism}

We consider a general Hamiltonian in QOC problem as 
\begin{equation}
H(t)=H_{d}+H_{c}(t)  \label{Eq1}
\end{equation}%
where $H_{d}$ is the drift Hamiltonian of the system. $H_{c}(t)$ is the
time-dependent Hamiltonian which is to be optimized to control the quantum
system to undergo a desired time evolution. The dynamics of the system
steered by the total Hamiltonian in Eq. (\ref{Eq1}) satisfied the Schr$\ddot{%
o}$dinger equation $|\dot{\psi}(t)\rangle =-iH(t)|\psi (t)\rangle $, with
the time evolution operator satisfying $\dot{U}(t)=-iH(t)U(t)$, or an
integral form $U(t)=\mathcal{T}e^{-i\int H(t)dt}$, where $\mathcal{T}$ is
the time-ordering operator. A generic form of the control Hamiltonian is $%
H_{c}(t)=H_{c}(\{\Omega _{j}(t)\},t)$, where $\{\Omega _{j}(t)\}$ are a set
of time-dependent control pulse to be optimized. In our examples presented
later, $\{\Omega _{j}(t)\}$ are chosen to be the envelopes of microwave
drives on qubits, where $j=1x,1y,2x,2y$, i.e. $1x$ means rotating around X
axis for qubit 1. For different QOC problems, expectation function could be
customized. A constrained optimization could be perform by combining penalty
functions into the expectation function with Lagrange multipliers \cite{shu2016frequency, plick2015physical}. In the specific examples discussed in this article, we study the
performance of a quantum gate at final time $t=T$ as an average over all
possible initial quantum states. It can be quantified by the average gate
fidelity~\cite{Fidelity} defined as 
\begin{equation}
f=\frac{1}{d(d+1)}(Tr(MM^{\dag })+|Tr(M)|^{2})  \label{Eq51}
\end{equation}%
where $M=U_{tar}U_{T}$, $d$ is the dimension of the quantum system.
Therefore, we use infidelity as the cost function to be minimized, i.e. $%
F=1-f$.

\subsection{Realistic requirements}

Traditional pulse optimization algorithms are confronted with many issues while
applying to realistic systems. We summarize various realistic issues below
and design the COCOA optimization to bridge the gap between numerical
optimization and experimental applications.

1. \textit{Pulse pre-distortion.} Pulse distortion as one of the major issues
could take place in the following process~\cite{rabiner1975theory}: 
i) Pulse generation with finite sampling rates,
which could be modeled as FIR filter. ii) Transmission of
signal, where a IIR transfer function could be used
to model the distortion. iii) Numerical
distortion when post-processing the optimal pulse for the purposes such as smoothening, which could be modelled as the post-optimization filter function. Commonly, the processed pulse is not optimal any more. Pulse pre-distortion with inverse filter function could be applied in experiments to compensate these effects\cite{chen2018metrology,rol2020time}. Since the filter functions are in frequency domain, it would be much more convenient for pre-distortion if the numerical output pulse is an
analytical function for frequency rather than in the PWC form. 

2. \textit{Pulse constraints:} Optimal pulses should satisfy various physical constraints, such as smoothness, finite bandwidth, bounded amplitude, starting and ending at some designated values, robustness to some randomness (e.g. noises), and so on. Post-processing the optimized pulses with constraints results in numerical pulse distortion mentioned above. So it is necessary to incorporate constraints into the optimization. However, efficiently adding pulse constraints to the optimization is challenging, because the gradient might be too complicated to compute. Traditionally, conditional expressions such as if-else paragraph could be added into the optimization solver. But this results in possible loopholes and low efficiency. So a better approach is still open for investigation. 

3. \textit{Pulse parameters.} Optimizing the pulse parameters in experiments
is necessary and could become exponentially strenuous when the number of
parameters grows. Therefore, it is desired to obtain optimal pulses with few
parameters for experimental tune-ups.

4. \textit{Analyticity of the pulse}. The analyticity is defined as explicit and analytical function or a definite summation of several analytical functions, which helps generating and tuning control pulses in experiments.

In this article, we will show how the COCOA algorithm is engineered to satisfy the above requirements efficiently. And we will test its performance via realistic optimization tasks. 

\section{COCOA Algorithm}

\label{section3}

\subsection{List of symbols}

\label{symbols}

\begin{tabular}{ll}
\hline\hline
Symbol & Meaning \\ \hline
$j$ & Index of control pulse $(1x, 1y, 2x, 2y)$ \\ 
$k$ & Time slice index $(1,\dots,N)$ \\ 
$\Omega [k] $ & Control amplitude in time slice $k$ \\ 
$\mathcal{C,M,E,F}$ & Transform function for each node: pulse \\ 
& constraint, bandwidth control, evolution, \\ 
& cost function \\ \hline
$\Omega ^{r}(t)$ & Control pulse at iteration $r$ \\ 
$N_{c}$ & Number of Fourier components kept in \\ 
& optimized pulse \\ 
$R$ & Number of iterations in optimization \\ 
$\alpha $ & Learning rate \\ 
$\epsilon _{0}$ & Tolerance of cost function \\ 
$\epsilon _{1}$ & Tolerance of gradient norm \\ 
$T$ & Total gate time \\ 
$S_{0}$ & Initial state \\ 
$S_{T}$ & State at time $t=T$ \\ 
$U_{T}$ & Evolution operator at time $t=T$ \\ 
$U_{tar}$ & Target operator \\ 
$\Omega_j ^{\ast }(t)$ & Optimized pulse of $j$-th control \\ \hline\hline
\end{tabular}

\subsection{Pseudo code}

The pseudo code of COCOA algorithm can be seen in Algorithm \ref{alg::COCOA}.

\begin{algorithm}
\caption{Complex and Optional Constraints Optimization (COCOA) for pulse engineering}
\label{alg::COCOA}
\begin{algorithmic}[1]
\Require
cost function: $F$; initial pulse sequence: $\Omega_j^0(t)$; $N_c$; $\alpha$; Iterations: $R$; $\epsilon_0$; $\epsilon_1$
\Ensure
optimized pulse: $\Omega_j^*(t)$
\State Discretization: $\Omega_j^0(t)\rightarrow \{\Omega_j^0 [k]\}$;
\Repeat
\State record computational graph for auto-differentiation;
\State pulse constrain: \{$\Omega_j [k]\} \rightarrow \{\bar{\Omega}_j [k]\} $;
\State bandwidth control: \{$\bar{\Omega}_j [k]\} \rightarrow \{\tilde{\Omega}_j [k]\}$;
\State evolution: $U_k = e^{- i \left( H_d + {\underset{j}{\sum}} \tilde{\Omega}_j [k] H_j [t_k] \right) \Delta t}, U_T = U_{N}U_{N-1}\ldots U_1$;
\State cost function: $F(U_T, S_T)$;
\State calculate gradient using auto-differentiation: $\{\frac{\partial f}{\partial \Omega_j [k]}\}$;
\Until{($F(U_T, S_T)<\epsilon_0$ or $\| \frac{\partial f}{\partial \{\Omega_j [k]\}}\|<\epsilon_1$)};
\end{algorithmic}
\end{algorithm}

\subsection{Algorithm settings}

\begin{figure*}[th]
\centering
\raisebox{0.0\height}{%
\includegraphics[width=15cm]{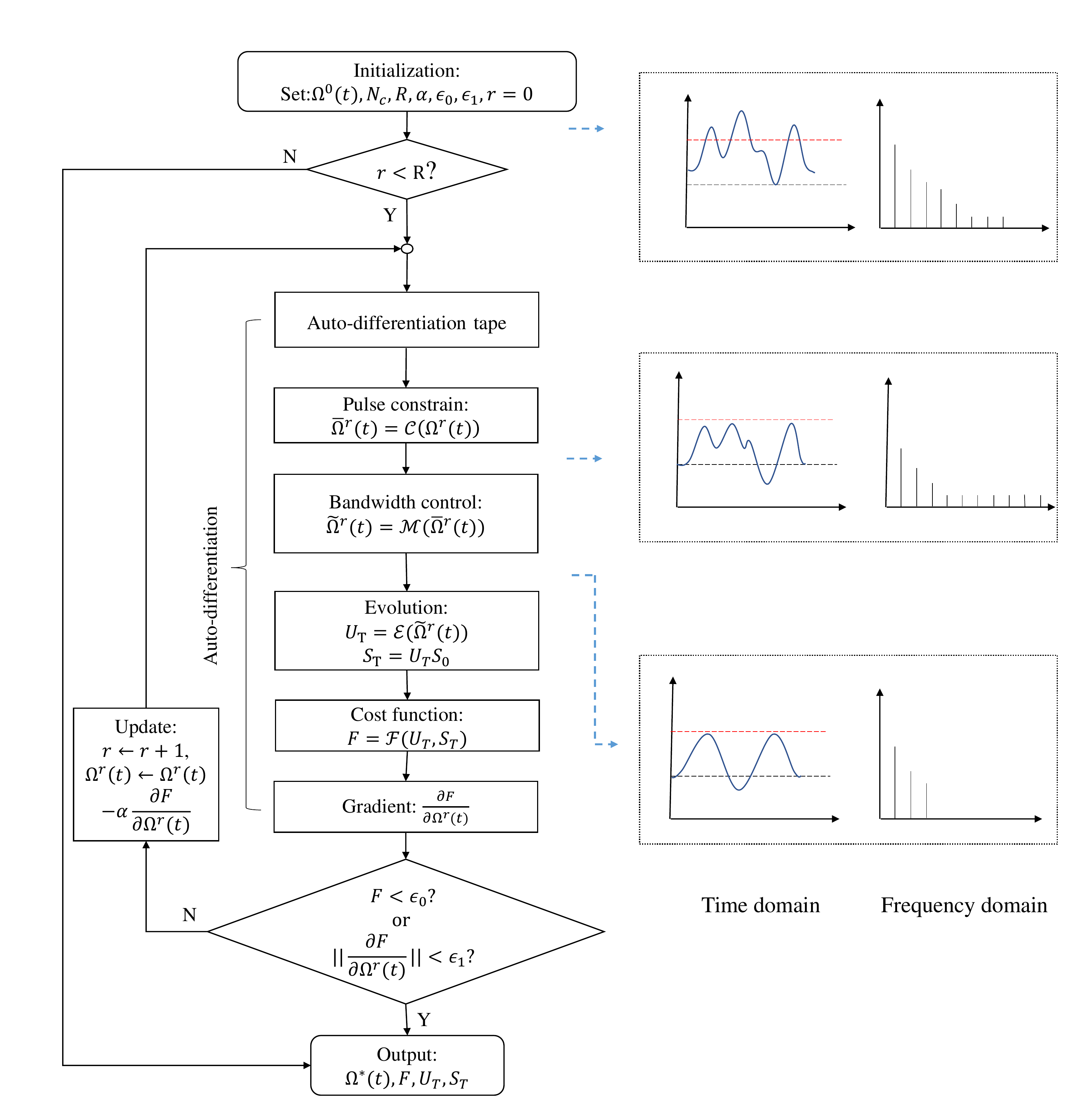}}
\caption{Flow diagram of COCOA. We use transform functions $\mathcal{C,M,E,F}$
to denote the 4 main nodes. Auto-differentiation tape is just a recorder
that records all the computing processes automatically by Tensorflow and will
be used for computing gradient. The figures on the left show the pulse's change in time and frequency domain after each specific node. }
\label{COCOA}
\end{figure*}

The process of COCOA is shown as a flowchart in Fig. %
\ref{COCOA}. The four main nodes in COCOA are pulse constraint, bandwidth
control, evolution, and cost function. Pulse constraints and bandwidth control
node are the core nodes, which always do a pretreatment
on the pulse before evolution, while the form of evolution node and cost function node depend on the problem and optimization task. In the
following, we elaborate each node and illustrate several unique features of CRAB.

\subsubsection{Ansatz for pulses}

The initial guess in COCOA can be chosen to be either a random guess or a specific
form based on prior knowledge. Then it is transformed to an function for
pulse parameters $\Omega _{j}(\vec{p},t)$, where $\vec{p}$ is the parameter
vector. The analytical form of this function could be arbitrary according to the need in
practice. For the consideration of limiting pulse bandwidth, without loss of
generality, we take chopped Fourier basis functions as 
\begin{equation}
\Omega _{j}(\vec{p},t)=a_{0}+\overset{N_{c}}{\underset{n=1}{\sum }}%
A_{jn}\cos (\omega _{jn}t)+B_{jn}\sin (\omega _{jn}t),
\end{equation}%
where pulse parameter set $\vec{p}$ is formed with Fourier expansion
parameters 
\begin{equation}
\vec{p}=\{A_{jn},B_{jn},a_{0}\}.
\end{equation}%
Analytical pulse functions solved from different theories could be exactly or
approximately transformed to the chopped Fourier basis for optimization,
such as Slepian pulses \cite{martinis2014fast}, SWIPHT
pulses \cite{swift1,swift2}, geometric pulses \cite{deng2021correcting}, and
so on. We will demonstrate this in Sec.\ref{swiftpulse}.  Note that filter functions for pulse pre-distortion could be directly
applied on the Fourier basis, which brings additional convenience to
experimental tune-ups. 

In the PWC optimization, $\Omega _{j}(\vec{p},t)$ is discretized to a $N$%
-length sequence with sampling frequency $f_{s}=N/T$, where $T$ is the total
gate time. 
\begin{equation}
\Omega _{j}(\vec{p},t)\overset{\text{ sample }}{\longrightarrow }\{\Omega
_{j}\left[ \vec{p},k\right] \},k=1,\ldots ,N.
\end{equation}%
For convenience, the discretized temporal sequence of the PWC ansatz is
denoted as $\{\Omega _{j}[k]\},k=1,...,N$. 

\subsubsection{Amplitude constraint}

A realistic system limits a maximal strength of the control field. Also, a
single control pulse starts and ends at zero. As a traditional way in
textbook ~\cite{Alessandro, GRAPE1}, this constraint enters the optimization
cost function by adding up with the control power defined as $J=\lambda
\int_{0}^{T}\Omega _{j}^{2}(t)dt$, where$\quad $the weight$\ \lambda >0$.
However, the fidelity of optimized pulse will be lower with this term
added. Furthermore, this way just gives a soft constraint on amplitude
maximum, which could be harmful when physical systems have a hard limit on
control amplitude. In our algorithm, a strong constraint to the pulse
amplitude is added by passing the control pulses through a sigmoid window
function, similar to GOAT \cite{GOAT}. 
\begin{align}
S_{down}(t,g)& =\frac{1}{1+e^{gt}} \\
S_{up}(t,g)& =1-S_{down}(t,g) \\
S_{amp}(\Omega_j[k],l,u)& =\left( 2S_{up}\left( \frac{\Omega_j[k]-\frac{u+l}{%
2}}{\frac{u-l}{2}},0.5\right) -1\right) \frac{u-l}{2}  \notag \\
& +\frac{u+l}{2},
\end{align}%
where $g$ is the ascent/descent gradient of the window function. $l$, $u$
are the lower and upper bound of the pulse's amplitude.

The total amplitude constraint transformation reads 
\begin{align}
& \mathcal{C}(\Omega _{j}[k],t,l,u,g,\Delta _{t})  \notag \\
& =S_{up}(\frac{t-\Delta t}{T},g)S_{down}(\frac{t-(T-\Delta t)}{T}%
),g)S_{amp}(\Omega _{j}[k],l,u),
\end{align}%
where $\Delta t$ is the width of the ascent(descent) edge. The first and the
second sigmoid function ensures zero amplitude at $t=0$ and $t=T$, and thus
satisfies the second physical constraint. The last one bounds the amplitude
to the $[l,u]$ range.

\subsubsection{Bandwidth control}

In this node, We modulate the control pulse to a bandwidth-limited one
in frequency domain. Firstly, we transform the pulse sequence from time
domain into frequency domain using discrete Fourier transform (DFT) 
\begin{equation}
X [n] = \overset{N}{\underset{k = 1}{\sum}} \Omega_j[k] e^{i (2 \pi n / N)k}.
\end{equation}
After DFT, We get a complex sequence $\{ X [n] \}, n = 1, \ldots, N$.
Assuming the upper cut-off frequency  is $f_{th}$, we can derive the maximal
Fourier component number $N_{c}$ 
\begin{equation}
N_{c} \leq \text{Int}(N \frac{f_{t h}}{f_{s}}) 
\end{equation}
where $\text{Int}(\cdot)$ indicates rounding down. Here, $N_{c}$ is a
hyper-parameter in our algorithm and it affects the pulse's simplicity,
smoothness and numerical accuracy. We will discuss it in details in Sec. \ref%
{section4}.

After DFT, the higher Fourier components over $N_{c}$, namely the complex
sequence elements from $X [N_{c} + 1]$ to $X [N - N_{c}]$ will be set
to zero, i.e. limiting the bandwidth. Then the complex sequence in frequency domain becomes

\begin{equation}
Y [n]=\left\{ 
\begin{array}{cll}
\tilde{X} [n], & n\in[1, N_{c}] \cup[N-N_{c}, N] &  \\ 
0, & n \in[N_{c} + 1, N-N_{c}] & 
\end{array}
\right.
\end{equation}

Then we apply the inverse transformation of DFT, called IDFT, to transform
the pulse back to time domain \ 
\begin{equation}
\tilde{\Omega}_{j}[k]=\frac{1}{N}\overset{N}{\underset{n=1}{\sum }}%
Y[n]e^{i(2\pi k/N)n}.
\end{equation}%
After IDFT, the pulse sequence $\{\tilde{\Omega}_{j}[t_{k}] \}$ is a smooth pulse
sequence with limited bandwidth. Its functional form in continuous time
domain is denoted as 
\begin{equation}
\tilde{\Omega}_{j}(t)= a_{0}+\sum_{n=1}^{N_{c}}A_{n}\cos (n%
\frac{2\pi f_{s}}{N}t+\phi _{n}) .  \label{Eq18}
\end{equation}%
It is worth noting that this is the functional form of our final optimized waveform, which is a finite Fourier basis function. More details about DFT and IDFT can be seen in appendix \ref{appendixA}.

\subsubsection{Evolution}

For the evolution node, the smooth, analytical and bandwidth-limited control
pulse, Eq. (\ref{Eq18}), obtained from the previous nodes is taken into the
dynamical equation to compute the time evolution. The choice of evolution equations,
such as master equation and Schr$\ddot{o}$dinger equation, depends on the
specific physical problem and optimization task. Here we consider a closed
quantum system and use the PWC approach for the time evolution. Note that this could be upgraded to other finite difference methods to obtain more precise solution of the Schr$\ddot{o}$%
dinger equation. Here, the evolution operator at time $t_{k}$ reads 
\begin{align}
U_{k}& =e^{-iH[t_{k}]\Delta t}  \notag \\
& =e^{-i\left( H_{d}+{\underset{j}{\sum }}\tilde{\Omega}%
_{j}[t_{k}]H_{j}[t_{k}]\right) \Delta t}.
\end{align}%
Then final evolution operator at time $t=T$ reads 
\begin{equation}
U_{T}=U_{N}\ldots U_{1}U_{1}.
\end{equation}%
The final state $S_{T}$ reads 
\begin{equation}
S_{T}=U_{T}S_{0}.
\end{equation}

\subsection{Pulse distortion in numerical process}

\label{distortion}

There are at least three steps of pulse distortion in a complete quantum
control task. First, the pulse distorted from the waveform optimized
numerically because of the finite sampling rate of the arbitrary waveform
generator (AWG)~\cite{lin2018high, raftery2017direct}. Second, the pulse
experience distortion during the transmission due to impedance mismatching
and other realistic filtering effects~\cite{rol2020time,chen2018metrology,baum2021experimental,hincks2015controlling}. Third, to force the optimized pulses satisfy physical constrains, pulse distortion is often induced when post-processing the output pulses from optimization iteration, such as adding filter functions to the output. 
As mentioned above, PWC algorithms, such as GRAPE and Krotov, generates
rough pulses. In order to smoothen the optimal pulses, low-pass filter such
as Gaussian filter ~\cite{ziegel1988filtering} could be applied to suppress
or cut off the high frequency components and limit the bandwidth, after
which the resultant waveform deforms and the fidelity is lowered from the
optimum. On the other hand, limiting pulse amplitude~\cite{GOAT} by
applying a constraint function to the optimization results in distortion
from the analytical form and induces additional high frequency
components.\ However, COCOA introduce all the pulse constraints into the
optimization before the DFT node and all the high frequency component will
be filtered, overcoming this kind of pulse distortion perfectly.
Additionally-customized constraints could be incorporated as well. This is
enabled with the use of auto-differentiation in Tensorflow, which is
discussed next. As a result, the optimized pulse is band-limited,
maximum-limited, starting/ending at ZERO while a definite analytical
waveform is guaranteed to output from the algorithm. This will be
illustrated in Sec. \ref{section4}

\subsection{Auto-differentiation}

\begin{figure*}[th]
\centering
\raisebox{0.0\height}{\includegraphics[width=1.6%
\columnwidth]{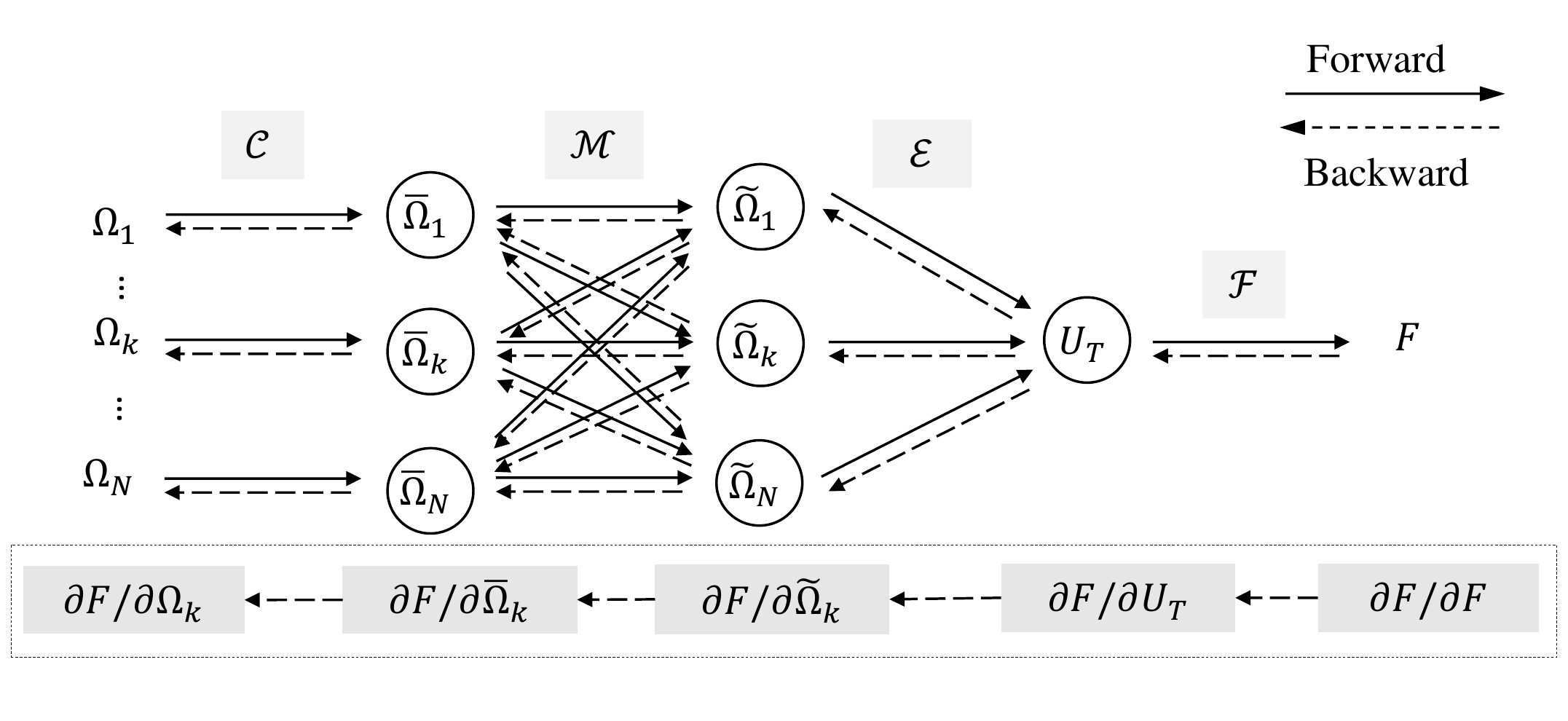}}
\caption{Computational graph of auto-differentiation. $\mathcal{C,M,E,F}$ denote four transform layers, corresponding to each node. ${\overline{\Omega}_k}, {\widetilde{\Omega}_k}, U_T, F$ are the output sequence after each transform layer. The solid line(dashed line)
is the forward propagation(backpropagation). Subscript $k$ denotes for the
index of the control sequence $\Omega$. The dashed box at the bottom shows the process of computing the gradient of $F$ for $\Omega_{k}$, using the chain rule.}
\label{AD}
\end{figure*}

Auto-differentiation (AD) method is widely used in machine learning, which is almost 
as accurate as symbolic differentiation \cite{Auto-differentiation}. There are two points of necessity that we choose AD: (1) AD obtains
the derivatives to all inputs in one back-propagation when AD works in the reverse mode. So it is much more efficient than manual and symbolic differentiation, and is more precise than numerical differentiation. (2) The complexity of derivatives induced by
the bandwidth control and the extraction of computational subspace places much
difficulties, such as expression swell problem, to manual and symbolic differentiation. The feasibility of auto-differentiation is demonstrated by the fact that each node of COCOA is derivable theoretically, so as the total transform function of inputs to cost function is $\mathbb{F}:\mathbb{R}%
^{n}\rightarrow \mathbb{R}$. This is suitable for the reverse mode of AD
because the dimension of inputs is larger than outputs. We explicitly elaborate the process of auto-differentiation in COCOA optimization in Fig. \ref{AD}. There are two processes
when using auto-differentiation in reverse mode. i) Forward-propagation: when computing
from $\{\Omega_j [k]\},k=1,...,N$ to $F$, it automatically constructs
a computational graph formed of nodes and edges, as shown in Fig. \ref{AD} (solid line). ii) Back-propagation: The gradients
of $F$ versus all inputs $\{\Omega_j [k]\},k=1,...,N$ are
calculated with the computational graph by using chain rule, as shown in Fig. \ref{AD} (dashed line). We note that all the derivatives of $F$ with respect to inputs $\{\Omega_j
[k]\},k=1,...,N$ are calculated in one back-propagation, which makes AD
more efficient than other methods. Our numerical simulation results in Sec. %
\ref{section4} will demonstrate the AD's efficiency in quantum
optimal control. For more details of AD, please refer to ~\cite%
{Auto-differentiation}.

\section{Applications on Superconducting Qubits}

\label{section4} To demonstrate the advantage of COCOA, we apply this
algorithm to tackle one of the most challenging control obstacles in the
up-to-date multi-qubit processors: the always-on couplings. Such issue lies
in many qubit systems such as superconducting qubits ~\cite%
{zhao2020high,ku2020suppression,deng2021correcting}, quantum dots~\cite%
{khaetskii2002electron, jacak2013quantum, maksym1990quantum}, NMR qubits~%
\cite{vandersypen2005nmr, wang2016single}, etc. Scaling-up qubit systems
tends to reduce the number of control degrees, which means taking out more
control fields out of the system. Losing either the control of qubit
frequencies or coupling strength brings more difficult in realizing good
quantum gates, especially in the systems such as fixed frequency qubits with
fixed couplings, tunable qubits with residual couplings, qubits with tunable
couplings with unwanted interaction and crosstalk ~\cite%
{sarovar2020detecting, ash2020experimental, murali2020software}.
Fortunately, the degree of control freedom on the pulse-shaping could be
further exploited with the help of COCOA. In this section, without loss of
generality, we consider two realistic models of multi-connected
superconducting qubits and apply COCOA to find optimal control pulses for
single-qubit and two-qubit gates for the always-coupled qubits.

\subsection{Model 1: Two transmon qubits coupled directly}

In X-mon (X-shaped transmon) arrays where qubits are coupled directly via a capacitance with a constant interaction $g$, such as Google's previous version of quantum computing chip Bristlecone~\cite%
{kelly2019operating} and other chip design with few control lines ~\cite{long2021universal}. As a simplified model, we consider two qubit
coupled directly and obtain the Hamiltonian as%
\begin{equation}
H_{0}=\underset{j=1,2}{\sum }\omega _{j}a_{j}^{\dag }a_{j}+\frac{\alpha _{j}%
}{2}a_{j}^{\dag }a_{j}^{\dag }a_{j}a_{j}+g_{12}(a_{1}^{\dag
}+a_{1})(a_{2}^{\dag }+a_{2})
\end{equation}%
This model is effectively valid for qubits coupled via tunable couplers ~%
\cite{arute2019quantum,deng2021correcting}. Here $\omega _{j}$ are the qubit
frequencies and $\alpha _{j}$ are the anharmonicities of transmon qubits, $%
j=1,2$. To implement single qubit operations, neighbour qubits are detuned
with $\Delta =\omega _{1}-\omega _{2}$ and the effective zz-coupling
strength is turned down with the rate $\frac{g_{12}^{2}}{\Delta } $~\cite%
{deng2021correcting, krantz2019quantum, kjaergaard2020superconducting,
wendin2017quantum}. $g_{12}$ is the capacitive coupling strength between two
qubits. $a^{\dag }(a)$ denotes the qubit creation (annihilation) operators. This
unwanted coupling gives rise to frequency splitting between $\left\vert
00\right\rangle \rightarrow \left\vert 01\right\rangle $ and $\left\vert
10\right\rangle \rightarrow \left\vert 11\right\rangle $, inducing gate
errors, as well as control crosstalk~\cite{deng2021correcting}. This could
be a more challenging issue when qubit frequencies are fixed~\cite%
{long2021universal}. Complex control pulses are proposed to resolved this
issue but finding appropriate pulses remains a difficulty~\cite%
{deng2021correcting, GRAPE1}. The microwave pulses are sent in to drive the
transmons via this operator%
\begin{equation}
H_{d}^{j}=a_{j}^{\dagger }e^{-i\omega _{d}t}+a_{j}e^{i\omega _{d}t}
\label{drive}
\end{equation}%
where $\omega_{d}$ is the driving frequency. The waveform $\Omega (t)$ is
applied to the drive and modulates the strength of the control pulse. Hence,
the total Hamiltonian reads 
\begin{equation}
H=H_{0}+\sum_{j}\Omega_{j}(t)H_{d}^{j}
\end{equation}%
The control field could be added to both qubit 1 and 2 simultaneously or
only on a single qubit 1 or 2.

To illustrate the properties of COCOA's solutions and demonstrate ths
advantage of the algorithm, we show some numerical examples of optimizing
quantum gates in realistic system by comparing different algorithms,
including COCOA, CRAB, and GRAPE. For a fair comparison, we use gradient
descent optimizer Adam~\cite{gradient_method} in all these algorithms, but
keep the rest steps the same as the original versions. Therefore, we denote
them as COCOA, GRAPE-like and CRAB-like in our results. In the simulation,
each transmon is truncated to a four-level system to better consider
leakage. The model parameter we used is similar to Ref.~\cite{qiu2021suppressing} as $\omega _{1}/2\pi =5.270$ GHz, $\omega
_{2}/2\pi =4.670$ GHz, $\alpha _{1}/2\pi =\alpha _{2}/2\pi =220$ MHz, $%
g_{12}/2\pi =25.4$ MHz. The initial pulses for all algorithms are identical and take the form as shown in Eq. (\ref{Eq18}).

Note that the coupling strength between the two qubit is at the order of $%
g/\omega  \simeq 10^{-2}$. The results for the weak coupling $g/\omega \leq 10^{-3}$ and
ultra-strong coupling regime $g/\omega \geq 10^{-1}$ are shown in appendix \ref{appendixc}, all of which demonstrating
the enhancement of COCOA in the search of optimal pulses.

\subsubsection{Single qubit X gate at the presence of interaction}

\label{singlequbit} The first gate is a single X rotation only on the second
qubit while remaining the state in the first qubit. The target evolution
operator of the two qubit system is 
\begin{equation}
U_{tar}=I\otimes \sigma _{x}  \label{28}
\end{equation}%
which include both qubit's dynamics in the computational subspace $span\{%
\widetilde{\left\vert 00\right\rangle}, \widetilde{\left\vert 01\right\rangle%
}, \widetilde{\left\vert 10\right\rangle}, \widetilde{\left\vert
11\right\rangle}\}$. The identity $I$ in the first qubit's subsystem meets
the requirement that the control field does net-zero operation to the first
qubit at the presence of crosstalk due to the coupling. Applying simple
pulses results in entanglement between this two qubits. However, COCOA can
efficiently find composite pulses to achieve the $U_{tar}$ evolution. Here
we set $\omega _{d}$ in resonance with transition $\widetilde{\left\vert
00\right\rangle} \rightarrow \widetilde{\left\vert 01\right\rangle} $, then
it is off-resonant with $\widetilde{\left\vert 10\right\rangle} \rightarrow 
\widetilde{\left\vert 11\right\rangle} $ ~\cite{deng2021correcting}. Other relevant parameters are $T=50$ns, $N=148$, $f_{s}/2\pi =N/T=2.96GHz$, $%
N_{c}=5$. The range of pulse amplitude is $[-30, 30]$ MHz$\times 2 \pi$.

\begin{figure*}[th]
\raisebox{0.0\height}{\includegraphics[width=2\columnwidth]{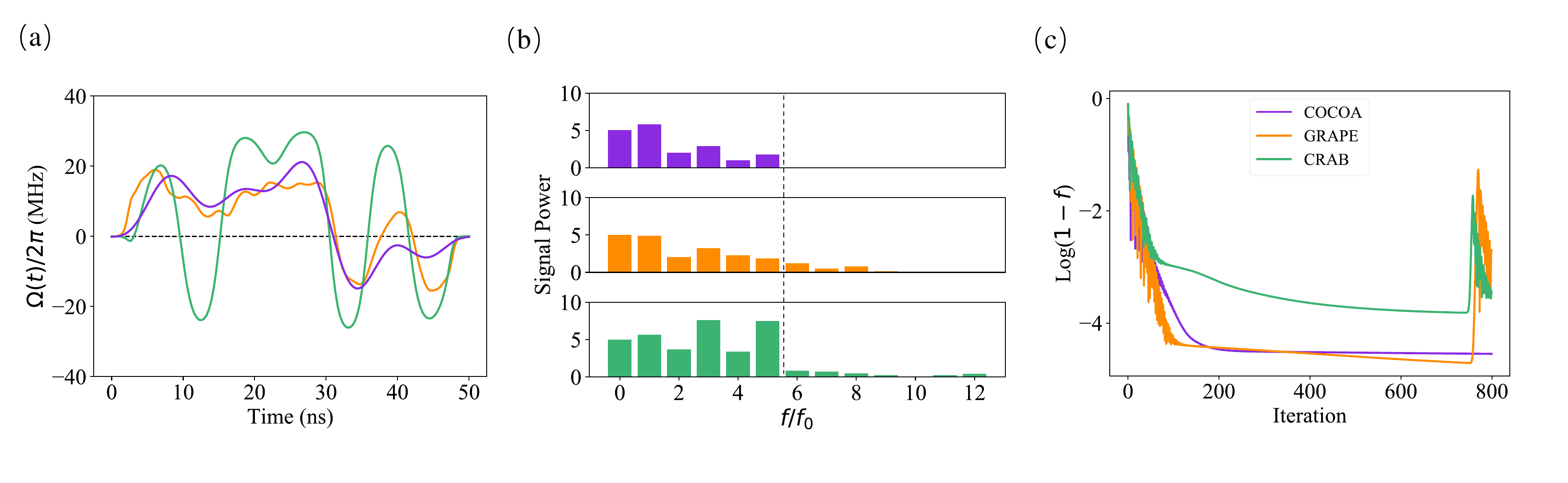}}
\caption{Algorithm comparison.(a) Optimized pulse with zero-point at $t = 0$ ns and $t = 50$ ns. Pulse amplitude is limited between $[-30, 30]$ MHz. (b) Frequency spectrum of each pulse. $f_0 = 1/T=0.02$ GHz. As we can see that
GRAPE(orange line) and CRAB(green line) have more high-frequency
components induced by pulse constraints. The dashed line shows the truncated components of the COCOA algorithm. (c) Gate infidelity changing versus iteration, showing the
efficiency of each algorithm. }
\label{res1}
\end{figure*}

We can conclude three advantages of COCOA according the comparison result in Fig. \ref{res1}: (1) Accuracy. As shown in Fig. \ref{res1} (a), COCOA achieves the gate infidelity below $%
10^{-4}$, which is the same order of magnitude with GRAPE-like's result and better than CRAB-like's. Even better results could be obtained by
enlarging $N_{c}$, as illustrated later in \ref{4.2}. (2) Smoothness and bandwidth. As shown in Fig. \ref{res1} (b), the optimal pulse obtained by COCOA
shows the best smoothness and limited pulse amplitude within $[-20,20]$ MHz$\times 2\pi$. From the frequency spectrum of optimized pulse, as shown in Fig. \ref{res1} (c), COCOA has limited bandwidth but the other's consist of high-frequency components. Interestingly, GRAPE-like's pulse shows a similar profile as COCOA,
with more high frequency components. Although CRAB-like's pulse looks smooth too, its amplitude is significantly stronger than the other two. (3) Analyticity. As promised, COCOA gives a definite analytical summation
form of the basis, all the high frequency components are filtered
completely. Specifically, COCOA outputs the pulse parameters for Eq. (\ref%
{Eq18}) with 
\begin{align*}
a_{0}& =5.066 \\
A_{1},\ldots ,A_{5}& =-11.66,-4.172,-5.753,2.140,3.497 \\
\phi _{1},\ldots ,\phi _{5}& =1.080,-3.385,6.104,-1.458,1.098
\end{align*}%
GRAPE-like and CRAB-like both produce uncontrollable high
frequency components and cannot obtain definite analytical expressions due to the ZERO starting and ending point constraint.

COCOA shows a great advantage here. Because the smoothness of control pulse
is rather important in many qubit systems where leakage and crosstalk errors
are significant. Limiting the bandwidth could also help reduce pulse
distortion throughout all the control steps. Moreover, the analytical
expression with definite summation of few chopped basis brings convenience
and simplicity to the experimental adjustments. It is necessary to point out
that, in principle, CRAB also uses analytical waveform with definite
summation of chopped random basis, but the pulse constrains during
optimization process induces numerical pulse distortion from the original analytical form, leading to higher frequency components, as shown in Fig. \ref{res1}(c).
Compared to GRAPE and CRAB, COCOA is so far the first algorithm producing
definite analytical pulses.

\subsubsection{Dual X gate at the presence of interaction}

\begin{figure}[th]
\raisebox{0.0\height}{\includegraphics[width=1%
\columnwidth]{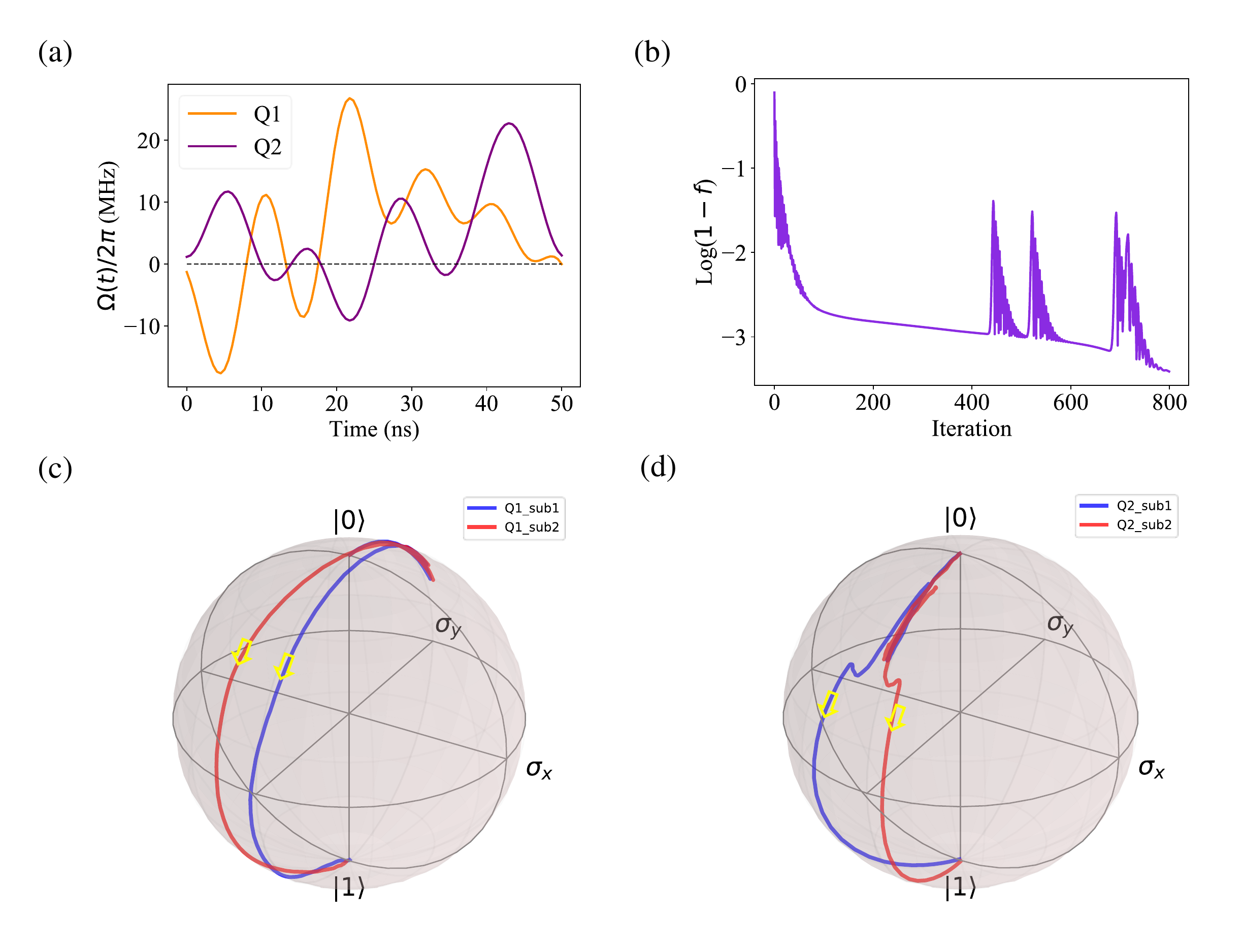}}
\caption{Dual X gate optimization using COCOA. (a) Optimized control pulse for each qubit with pulse amplitude limitation $[-30, 30]$ MHz and ZERO starting and ending point. (b) Gate infidelity changes versus iteration, indicating the efficiency of the algorithm.  (c) Evolution trajectories of qubit 1, driven by the darkorange pulse in (b). Blue(red) line shows the evolution in on-resonant (off-resonant) subspace span $\{\widetilde{\left\vert 00\right\rangle}, \widetilde{\left\vert 01\right\rangle} \}$ ($\{\widetilde{\left\vert 10\right\rangle}, \widetilde{\left\vert 11\right\rangle} \}$). (d) Evolution trajectories of qubit 2, driven by the purple pulse in (b). Blue and red line have the same meaning in (c)}
\label{xxgateres}
\end{figure}

Simultaneously driving coupled qubits is challenging task even in tunable qubit system where the interaction could be tuned on and off approximately. Crosstalk of control signals and zz-interaction cause significant drop of gate error compared to individual driving case. In order to implement a high fidelity dual (simultaneous) X gate on both nearest-coupled coupled qubits, we apply COCOA to find the pulses to drive them at the same time. The Dual X gate simultaneously flip the states of the first and second
qubits. The target evolution operator of the two qubit system is
\begin{equation}
U_{tar}=\sigma _{x} \otimes \sigma _{x}
\end{equation}%
Both qubits should be driven simultaneously and the drive term follows the same form of Eq. \ref{drive}. Hence, the total Hamiltonian reads 
\begin{equation}
H=H_{0}+\Omega _{1}(t)H_{d}^{1}+\Omega _{2}(t)H_{d}^{2}
\end{equation}%
Here we use COCOA algorithm to find the optimized driving pulse of dual X gate and we  set $T=50$ns$,N=200,N_{c}=5,f_{s}=N/T=4$ GHz. As we can see in Fig. %
\ref{xxgateres}, the optimized gate fidelity is greater than $99.9\%$ and the pulse
parameter for each drive are given as following:

Pulse parameters for Q1: 
\begin{align*}
a_{0} & = 5.091 \\
A_{1}, \ldots, A_{5} & = -10.46, 2.971, 3.033, 6.725, -6.124, \\
\phi_{1}, \ldots, \phi_{5} & = 5.858, 1.547, 3.085, 1.458, -3.650
\end{align*}

Pulse parameter for Q2: 
\begin{align*}
a_{0} & = 5.078 \\
A_{1}, \ldots, A_{5} & = 7.790, 0.8855, 4.857, -6.030, 1.787 \\
\phi_{1}, \ldots, \phi_{5} & = 0.7156, 1.714, 2.507, 0.8538, 2.292
\end{align*}

By analyzing the driving pulse and its corresponding Bloch trajectory, we
found that the negative part of the driving pulse eliminates the detuning of
the off-resonance subspace with respect to on-resonance subspace, namely the zz-coupling.

\subsubsection{Optimizing $N_{c}$}

\label{4.2}

The key parameter $N_{c}$, i.e. the number of Fourier components, determines the smoothness and bandwidth of the pulse, as well as the number of optimizing parameters, which increases at a scaling rate of $2N_{c}$. Consequently, the choice of $N_c$ affects the optimization efficiency and accuracy.

\begin{figure}[th]
\raisebox{0.0\height}{\includegraphics[width=1\columnwidth]{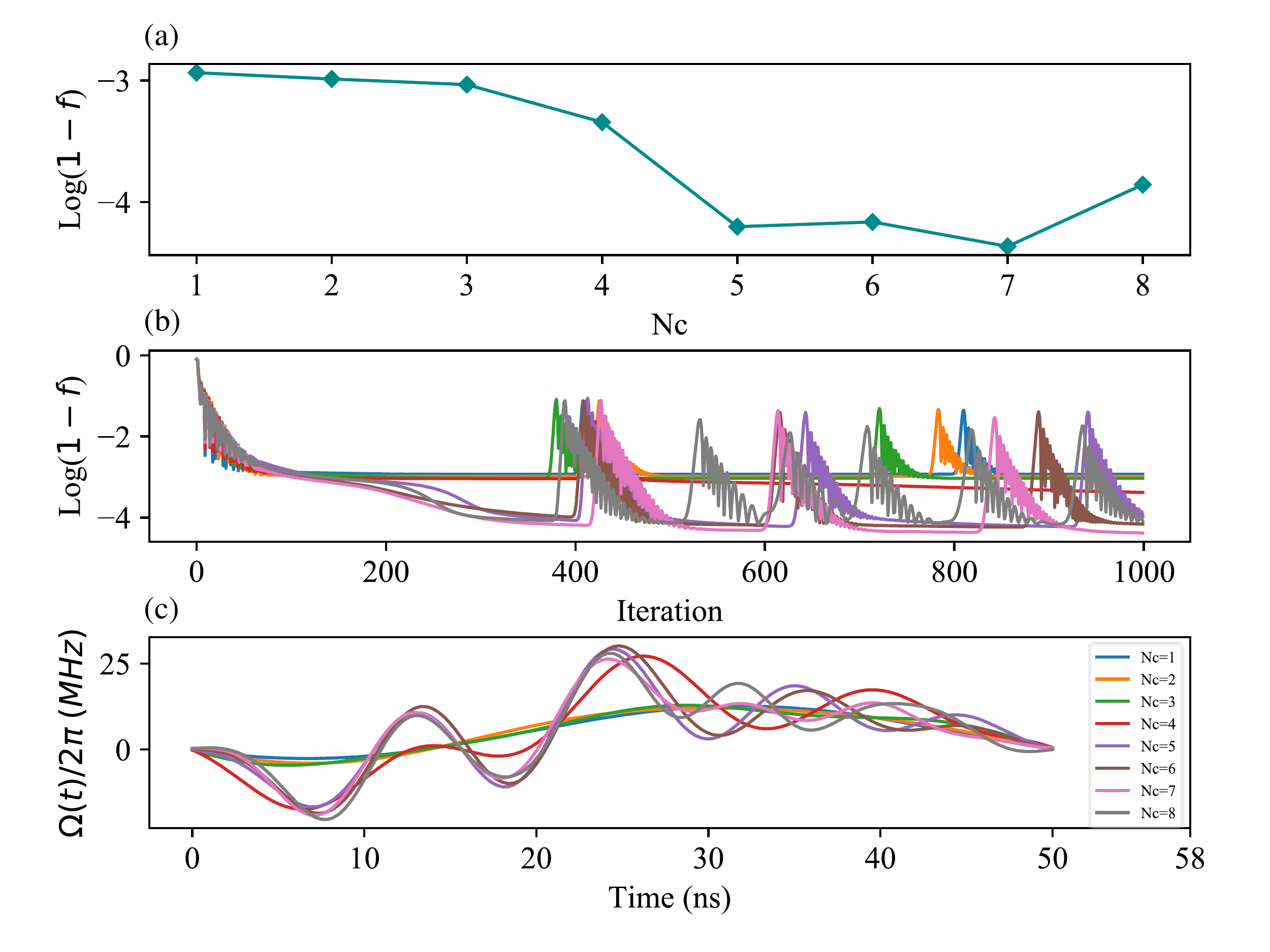}}
\caption{Explore how $N_c$ affects the accuracy, efficiency, and smoothness of our algorithm. (a) Best infidelity that COCOA can reach with each $N_c$. (b) Infidelity changes versus iterations for different $N_c$, indicating the efficiency. (c) Optimized pulse under different $N_c$. }
\label{Fig_Nc}
\end{figure}

We take the previous case in Sec. \ref{singlequbit} as an example to study
how $N_{c}$ affects the optimization, where we only tune $N_{c}$\ while fixing all
the other parameters. As shown in Fig. \ref{Fig_Nc} (a), gate infidelity is
improved by one order of magnitude when $N_{c}$ increases from $1$ to $5$,
and doesn't get much improved beyond $5$. Fig. \ref{Fig_Nc} (b) and (c) demonstrate the convergence behavior and pulse shape of $N_{c}=1, 2,..., 8$. From these simulation
results, we make an observation that $N_{c}=5$ is the best choice based on
the consideration of the trade-off between number of parameters and
optimization accuracy.

Theoretically, if there is no bandwidth limit, the COCOA algorithm can
approach to GRAPE algorithm when $N_{c}$ reaches its maximum: $%
N_{c}^{max}=\text{Int}\left( \frac{N-1}{2}\right) $, where $\text{Int}(\cdot )$ means
rounding down.

\subsection{Model 2: Two transmon qubits coupled via a Cavity}

In another widely used architecture, superconducting qubits are coupled via
superconducting cavities, such as one dimensional transmission line
resonators~\cite{majer2007coupling, gong2019genuine, chow2012universal}, with the Hamiltonian 
\begin{equation}
H_{0}=\underset{j=1,2}{\sum }\omega _{j}a_{j}^{\dag }a_{j}+\frac{\alpha _{j}%
}{2}a_{j}^{\dag }a_{j}^{\dag }a_{j}a_{j}+g_{cj}(b^{\dag }a_{j}+ba_{j}^{\dag
})+\omega _{c}b^{\dag }b
\end{equation}%
where $\omega _{c}$ are the frequency of the cavity. $g_{cj}$ is coupling
strength between resonator and the $j$-th qubit. $b(b^{\dag })$ is
annihilation(creation) operator of resonator. Other parameters have the same
meaning as in model 1. We take $\omega _{1}/2\pi =6.2$GHz, $%
\omega _{2}/2\pi =6.8$GHz, $\omega _{c}/2\pi =7.15$GHz, $\alpha _{1,2}/2\pi
=350$MHz, $g_{1,2}/2\pi =250$MHz, which are used in Ref.~\cite{model2}. The drive Hamiltonian $H_{d}
$ has the same form as Eq. \ref{drive}.

Since the cavity behaves as merely a larger scale fixed-coupler between two
qubits, control pulses for single-qubit gates could be obtained similarly as
previous discussion. Detailed numerical results of single qubit X gate could be found in appendix %
\ref{appendixd}. Here we demonstrate an optimization of two-qubit entangling
gate for this model.

\subsubsection{Optimizing CNOT gate based on SWIPHT protocal}

\label{swiftpulse} It's worth to point out that COCOA can fully utilize the
prior knowledge of analytical methods and obtains completely analytical
optimal pulses via local optimization around an analytically-given pulse.  To
demonstrate this, we start from a CNOT gate implementation using SWIPHT
protocol (speeding up wave forms by inducing phases to harmful transitions) 
\cite{model2,swift1,swift2,long2021universal}. The given analytical form of
the pulse is%
\begin{equation}
\Omega (t)=\frac{\ddot{\chi}}{2\sqrt{\frac{\delta ^{2}}{4}-\dot{\chi}^{2}}}-%
\sqrt{\frac{\delta ^{2}}{4}-\dot{\chi}^{2}}\cot (2\chi )  \label{Eq30}
\end{equation}%
where $\chi (t)=\frac{A}{T^{8}}t^{4}(T-t)^{4}+\frac{\pi }{4}$, $A=138.9$, $%
T=5.87/\left\vert \delta \right\vert $. $\delta =\omega _{\widetilde{10}%
\rightarrow \widetilde{11}}-\omega _{\widetilde{00}\rightarrow \widetilde{01}%
}$ is the detuning between the target and harmful transition in the
computational subspace $span\{\widetilde{\left\vert 01\right\rangle }$, $%
\widetilde{\left\vert 01\right\rangle }$, $\widetilde{\left\vert
10\right\rangle }$, $\widetilde{\left\vert 11\right\rangle }\}$, where the
first qubit is the control qubit and the second one is the target qubit.

\begin{figure}[th]
\raisebox{0.0\height}{\includegraphics[width=1\columnwidth]{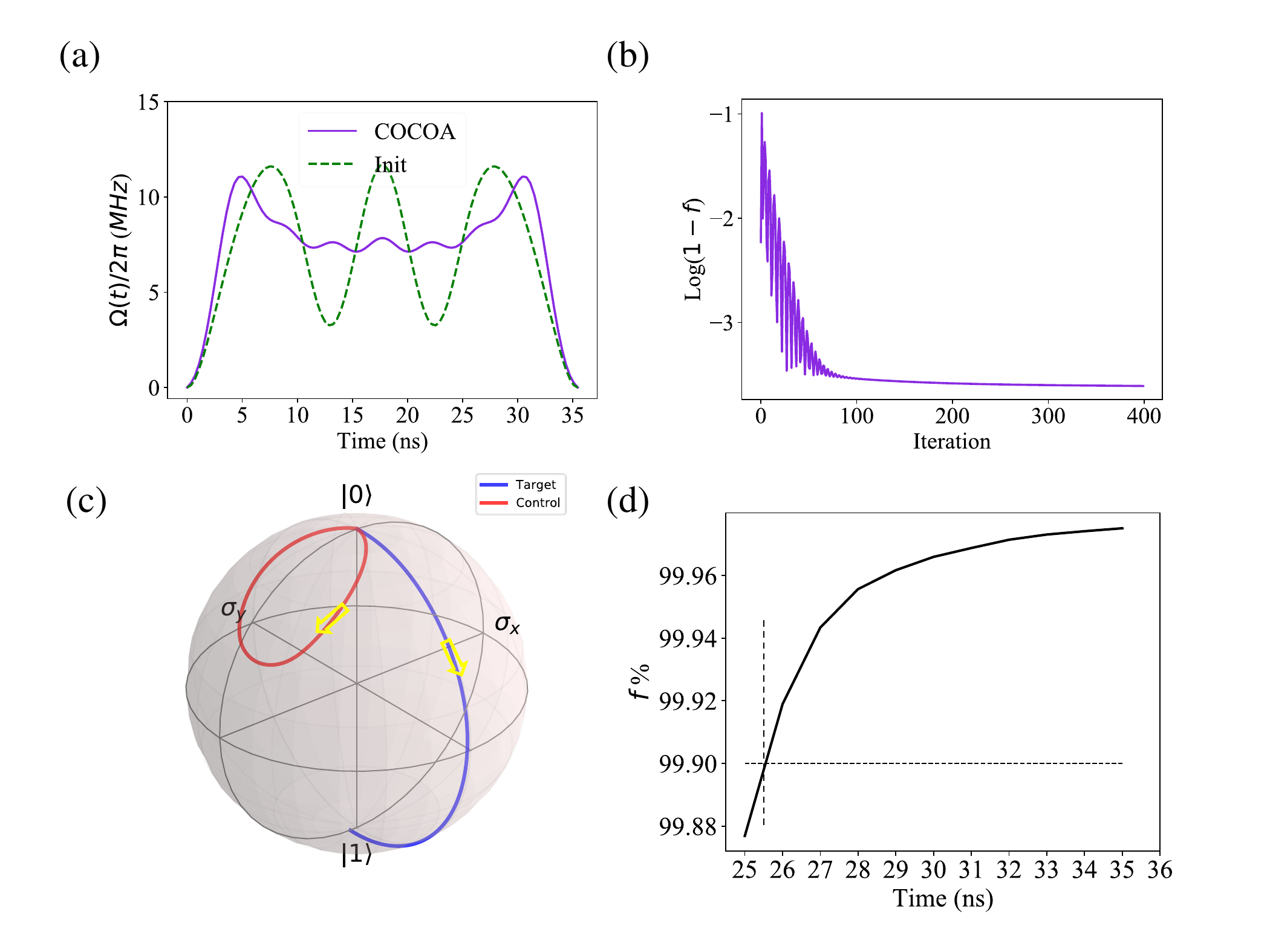}}
\caption{CNOT gate optimization using COCOA.  (a) Pulse comparison between the initial SWIFT CNOT pulse (green-dashed line) in Ref. ~%
\protect\cite{model2} and our optimized CNOT pulse (purple-solid line). (b) Gate infidelity changes
versus iteration, indicating the efficiency of the algorithm. (c)
Evolution trajectory on Bloch sphere driven by our optimized pulse in (b).
The blue and red lines show the target qubit's evolution trajectory in the
two subspaces $span\{\protect\widetilde{\left\vert 00\right\rangle },\protect%
\widetilde{\left\vert 01\right\rangle }\}\ $and $span\{\protect\widetilde{%
\left\vert 10\right\rangle },\protect\widetilde{\left\vert 11\right\rangle }%
\}$. (d) The gate speed limit of CNOT.}
\label{CNOTRes}
\end{figure}
The CNOT operator generated with a single microwave control could be
expressed as this general form%
\begin{equation}
U_{\text {tar }}=\left(\sigma_{x} \oplus I\right) \prod_{i=1,2; j=x, y, z} R_{i j}\left(\theta_{i j}\right)
\end{equation}
where $R_{ij}(\theta _{ij})$ are single qubit rotations with arbitrary
angles for optimization. This $U_{tar}$ is equivalent to a standard CNOT $%
\sigma _{x}\oplus I$ up to some local phases. We set drive frequency $\omega _{d}=\omega _{\widetilde{00}%
\rightarrow \widetilde{01}}$, and $N_{c}=9$. Fig. \ref{CNOTRes} (a) shows
that the local optimization converges very fast with prior knowledge of
optimal pulse, which demonstrates the advantage of numerical optimization
based on analytically-optimal pulses. Fig. \ref{CNOTRes} (b) shows that the
optimized CNOT pulses is transformed but still maintains similar shape as
the initial analytically-optimal pulse. We note that the optimized pulse,
show in Fig. \ref{CNOTRes} (b), can be generated more accurately with AWG
device due to its limited bandwidth that the initial pulse doesn't prosess.
With the same evolution time $T=35.4$ ns, we finally obtained the optimized
driving pulse shown in Fig. \ref{CNOTRes} (b) with complete pulse
parameters (in the unit of MHz) in Eq. \ref{Eq18} 
\begin{align*}
a_{0}& =7.416 \\
A_{1},\ldots ,A_{9}& =-0.818,-2.05,-2.27,-1.50-0.807, \\
& -0.2020.0287,0.325,0.00291 \\
\phi _{1},\ldots ,\phi _{9}& =0,\pi ,\pi ,0,0,0,0.000138,0,0.00198
\end{align*}%
The speed of this analytical CNOT gate is limited by $\frac{\delta ^{2}}{4}-%
\dot{\chi}^{2}>0$ derived from Eq. \ref{Eq30}. Hence, $T>0.02975\frac{A}{%
\delta }$, and we have $T_{min}=24.95$ ns when $A=138$ and $\delta =26.4$ MHZ%
$\times 2\pi $. The behavior of the optimal CNOT gate approaching the speed
limit is shown in Fig. \ref{CNOTRes} (c). Here, we start from $T=25$ ns and
increase by $1$ ns each step to observe the change of the optimal fidelity
with the gate time. Finally, we obtain the gate fidelity exceeds $99.9$\%
when $T\geqslant 26$ ns, which is a significantly improvement from the
initial CNOT gate time 35 ns using SWIFT theory. Here we note two important
tricks when using COCOA for a local optimization: 1. To obtain fast
convergence in this local optimization scenario, see Fig. \ref{CNOTRes} (a), a small learning rate is
favored, which is taken to be $0.001$; 2. The SGD
optimizer is preferred to Adam optimizer, since Adam is more suitable for
broader search range due to its momentum factor ~\cite{gradient_method}.

\section{Conclusion}

\label{section5} In this paper, we have developed a novel algorithm
to optimize smooth quantum control pulses constraints,
which could be very complex, highly nonlinear, sub/super-differentiable approximations
and optional. In the particular
demonstration examples, we limit the pulse amplitude, pulse bandwidth, and the number
of pulse parameters. Doing so makes this algorithm
involve complicated computation for the differentiation of expectation
function versus optimizing parameters. We resolve this issue using
auto-differentiation powered by Tensorflow. Therefore, this algorithm can be
straightforwardly extended to larger quantum systems with even more
complicated calculations of gradients. We have demonstrated the advantages
of the proposed algorithm by applying it to realistic superconducting qubit
models with always-on ZZ-interaction, and achieve optimal smooth pulses to
implement single-qubit gates and two-qubit gates. Comparing to GRAPE and
CRAB algorithms, we obtain higher gate fidelities and better optimization
efficiency. We have shown that COCOA could be applied to the optimization
scenarios either with or without good prior knowledge by simply switching
the optimizers, and obtain high-fidelity gates for both cases. 

We summarize COCOA's advantages as follows: 1. COCOA outputs optimal pulses with definite
analytical expression. 2. The optimal pulses have manually limited bandwidth
and amplitude. 3. This algorithm is more compatible with complex and
flexible pulse constraints, without the induced pulse distortion in
numerical optimization. 4. The auto-differentiation assisted by Tensorflow
enables efficient and easy calculation of gradients and the ability to handle
complex computing processes and complex models. 5. COCOA can speed up the
optimization by locally searching the optimal pulses based on certain prior
knowledge. Using COCOA, we completely resolve the challenging problem, to
implement either individually or simultaneously a single qubit X-gate in a
strongly ZZ-interacting two-qubit system \cite{long2021universal,
deng2021correcting, ash2020experimental}. In conclusion, COCOA optimization
is friendly to realistic quantum control tasks, easy to be customized, and
easy to be extended to larger quantum systems.

In conclusion, COCOA provides a versatile, highly functional and efficient platform to add physical constraints into quantum optimal control tasks. Following the line of COCOA, more work could be pursued in a near future to resolve the realistic QOC issues. For example, pulse pre-distortion could be effectively performed by adding the
definite transfer function of control lines and pulse generators into 
optimization process; Different analytical forms with fewer pulse
parameters could be investigated using the proposed algorithm, so that the
numerical approach could better meet the experimental needs. 

\acknowledgments
This work was supported by the Key-Area Research and Development Program of Guang-Dong Province (Grant No. 2018B030326001), the National Natural Science Foundation of China (U1801661), the Guangdong Innovative and Entrepreneurial Research Team Program (2016ZT06D348), the Guangdong Provincial Key Laboratory (Grant No.2019B121203002), the Natural Science Foundation of Guangdong Province (2017B030308003), and the Science, Technology and Innovation Commission of Shenzhen Municipality (JCYJ20170412152620376, KYTDPT20181011104202253), and the NSF of Beijing (Grants No. Z190012), Shenzhen Science and Technology Program (KQTD20200820113010023).

\section*{Author contributions}
YS and XHD conceived the project, designed the algorithm, performed calculations, and wrote the manuscript. YS did all the coding. JL gave some important suggestions on the algorithm and help design the flowchart. YJH helped with refining the algorithm. QG assisted with GRAPE optimization. XHD oversaw the project.

\begin{appendix}

\begin{figure*}[bh]
\raisebox{0.0\height}{\includegraphics[width=2\columnwidth]{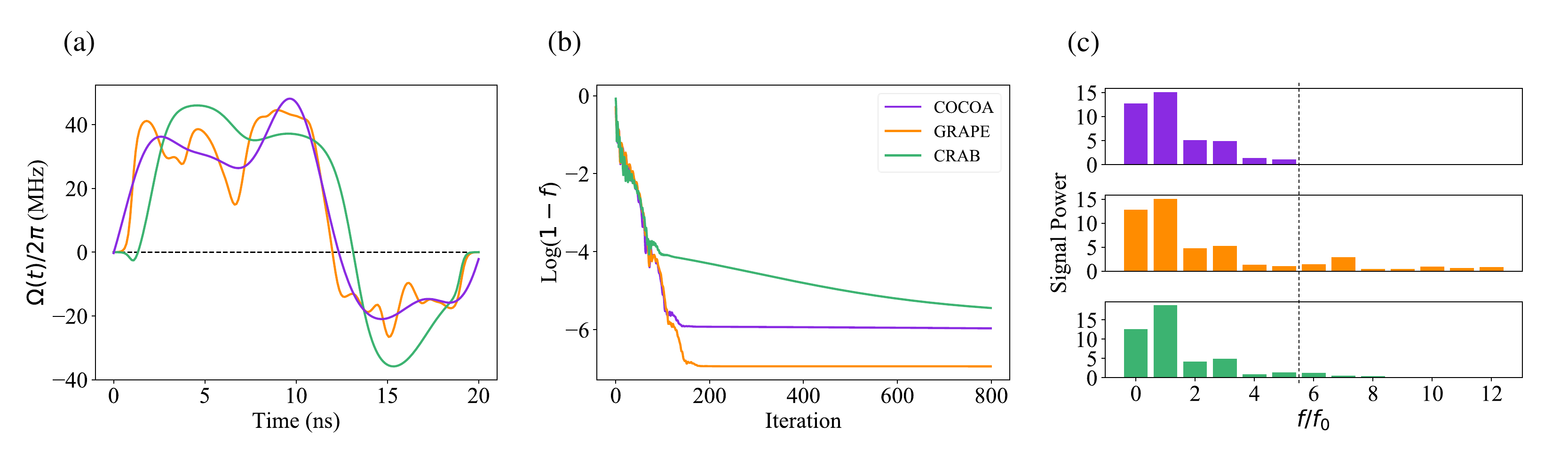}}
\caption{Algorithm comparison. $g=1$MHz$\times 2\pi$. (a) Optimized pulse with zero-point at $t = 0$ ns and $t = 20$ ns. Pulse amplitude is limited between $[-40, 40]$ MHz.(b) Gate infidelity changing versus iteration, showing the efficiency of each algorithm. (c)
Frequency spectrum of each pulse. $f_0 = 1/T=0.05$ GHz. As we can see that
GRAPE(orange line) and CRAB(green line) have more high-frequency
components induced by pulse constraints. The dashed line shows the truncated components of COCOA algorithm.}
\label{g=0.001}
\end{figure*}

\begin{figure*}[h]
\raisebox{0.0\height}{\includegraphics[width=2\columnwidth]{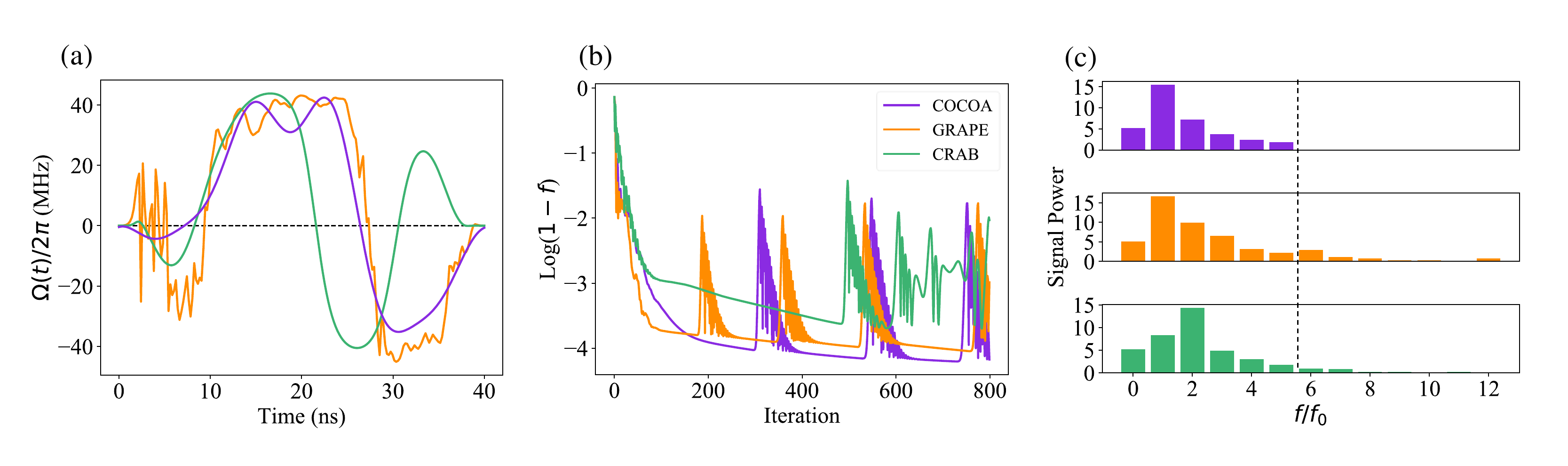}}
\caption{Algorithm comparison. $g=100$MHz$\times 2\pi$ (a) Optimized pulse with zero-point at $t = 0$%
ns and $t = 20$ns. The pulse amplitude is limited between $[-40, 40]$ MHz. (b) Gate infidelity changing versus iteration, showing the efficiency of each algorithm.  (c)
Frequency spectrum of each pulse. $f_0 = 1/T=0.025$ GHz. As we can see that
GRAPE (orange line) and CRAB (green line) have more high-frequency
components induced by pulse constraints. The dashed line shows the truncated components of COCOA algorithm.}
\label{g=0.1}%
\end{figure*}

\begin{figure*}[h]
\raisebox{0.0\height}{\includegraphics[width=2\columnwidth]{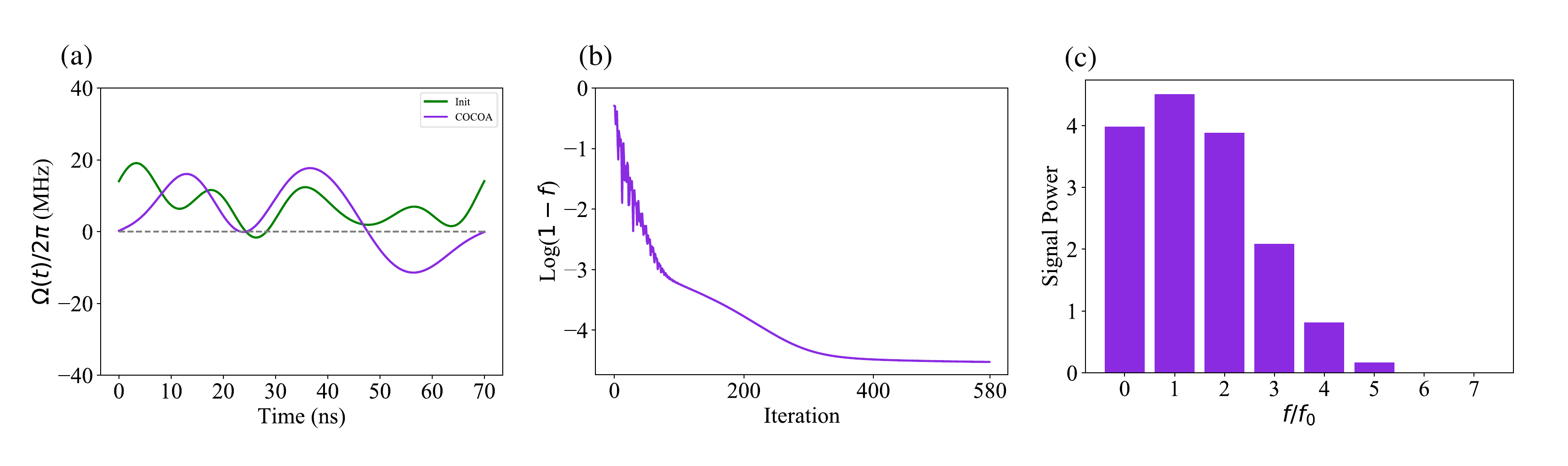}}
\caption{ Single X gate for model 2. (a) Optimized pulse with zero-point at $t
= 0$ ns and $t = 70$ ns. Pulse amplitude is limited between $[-20, 20]$ (unit:MHz). The purple line is the optimized pulse with COCOA algorithm and the green line is the initial Fourier pulse. (b) Gate infidelity changing versus iteration, showing the
efficiency of the algorithm.  (c) Frequency spectrum of COCOA pulse with limited-bandwidth($N_c=5$). $f_0=\frac{1}{T}=0.0143$GHz.}
\label{model2pigate}%
\end{figure*}

\section{Low Pass Filtering}
\label{appendixA}
\quad\textit{DFT:} Given a discretized pulse sequence in time domain $\Omega_j[k], k
= 1 \ldots N$. Note that the pulse here can be any form or without any
form. After DFT we can get a complex sequence which contains amplitude
and phase information.
\begin{equation}
X [n] = \overset{N}{\underset{k = 1}{\sum}} \Omega_j[k] e^{i (2 \pi n / N)k}.
\end{equation}

We can deduce that
\begin{align}
X[N-n]  &  = \sum_{k=1}^{N} \Omega_j[k] e^{-j(2 \pi/ N)(N-n) k}\\
&  = \sum_{k=1}^{N} \Omega_j[k] e^{j(-2 k \pi+(2 \pi/ N) k n)}\\
&  = \sum_{k=1}^{N} \Omega_j[k] e^{j((2 \pi/ N) k n)}\\
&  = X^{*}[n].
\end{align}
This is an important characteristic of the complex sequence.

\textit{IDFT:}%

\begin{equation}
\tilde{\Omega}_{j}[k]=\frac{1}{N}\overset{N}{\underset{n=1}{\sum }}%
Y[n]e^{i(2\pi k/N)n}.
\end{equation}

Assuming $N$ is oven number, We can deduce that
\begin{align}
N\times \Omega_{j}[k]  &  =\sum_{n=1}^{N}X[n]e^{j(2\pi/N)kn}\\
&  =X[0]+X[1]e^{j(2n\pi/N)}+X[2]e^{j(2(2n\pi/N))}+\cdots\nonumber\\
&  +X[N/2-1]e^{j\left(  \frac{2n\pi}{N}(N/2-1)\right)  } +X[N/2]\nonumber\\
&  +X[N/2+1]e^{j\left(  \frac{2n\pi}{N}(N/2+1)\right)  }+\cdots\nonumber\\
&  +X[N-2]e^{j\left(  \frac{2n\pi}{N}(N-2)\right)  }+X[N-1]e^{j\left(
\frac{2n\pi}{N}(N-1)\right)  }.
\end{align}

We set $X [n] = a_{n} + i b_{n}$. Considering sum of term $X[n]$ and $X[N-n]$,
by utilizing the conjugate property, we can deduce
\begin{align}
\left(  a_{n}+j b_{n}\right)  e^{j(2 \pi/ N) k n}+\left(  a_{n}-j
b_{k}\right)  e^{j(2 \pi/ N)\left(  N-k\right)  n}\nonumber\\
= A_{n} \cos\left(  \frac{2 k n \pi}{N}+\phi_{n}\right)
\end{align}
where $A_{n}=2 \sqrt{a_{n}^{2}+b_{n}^{2}}$ and $\tan\left(  \phi_{n}\right)
=\frac{b_{n}}{a_{n}}$. Then the total expression of $\Omega_{j}[k]$ in time domain reads
\begin{align}
N \times \Omega_{j}[k]  &  = a_{0} + A_{1} \cos(n 2 \pi/ N + \phi
_{1})\nonumber\\
&  + A_{2} \cos((2 n) 2 \pi/ N + \phi_{2}) + \cdots\nonumber\\
&  + A_{\frac{N}{2} - 1} \cos\left(  ((N / 2 - 1) n) 2 \pi/ N +\phi_{\frac
{N}{2} - 1} \right)  .
\end{align}

Assume sample frequency is $f_{s}$ we have
\begin{equation}
\Omega_{j}[k]  = \Omega_{j}[k T_s] = \Omega_{j} (k / f_{s}).
\end{equation}

Replace $k$ with $f_{s} t$
\begin{align}
N \times \Omega_{j}(t) &  = a_{0} + A_{1} \cos(f_{s} 2 \pi t / N + \phi
_{1})\nonumber\\
&  +A_{2} \cos((2 f_{s}) 2 \pi t / N + \phi_{2})+ \cdots\nonumber\\
&  + A_{\frac{N}{2} - 1} \cos\left(  ((N / 2 - 1) f_{s}) 2 \pi t / N +
\phi_{\frac{N}{2} - 1} \right)  .
\end{align}

And we can deduce the frequency at $k_{c}$ is $2 \pi k_{c} f_{s} / N = 2 \pi
f$.

\section{Weak and strong coupling strength} \label{appendixc}
This appendix demonstrates that COCOA algorithm can still be efficient in different coupling strength. In Fig. \ref{g=0.001}, we show the algorithm comparison with $g=0.001$GHz$\times 2 \pi$.
Fig. \ref{g=0.1} shows the result of $g=0.1$GHz$ \times 2 \pi$

\section{single qubit X gate with always on interaction for Model 2 }
\label{appendixd}
Here we show the result of single qubit X gate in Model 2 using COCOA algorithm to show its power and compatibility to multi-qubit system. The result is show in Fig. \ref{model2pigate}.

\end{appendix}


\end{document}